\documentclass[conference,final]{IEEEtran}
\makeatletter
\def\ps@headings{%
\def\@oddhead{\mbox{}\scriptsize\rightmark \hfil \thepage}%
\def\@evenhead{\scriptsize\thepage \hfil \leftmark\mbox{}}%
\def\@oddfoot{}%
\def\@evenfoot{}}
\makeatother 
\usepackage{amsmath}
\usepackage{amssymb}
\usepackage{graphicx,color}

\def\QED{\mbox{\rule[0pt]{1.5ex}{1.5ex}}}
\def\endproof{\hspace*{\fill}~\QED\par\endtrivlist\unskip}
\newtheorem{theorem}{Theorem}

\newtheorem{lemma}{Lemma}

\newtheorem{proposition}{Proposition}

\IEEEoverridecommandlockouts

\begin{document}

\title{Delay-Guaranteed Cross-Layer Scheduling in Multi-Hop Wireless Networks}

\author{\IEEEauthorblockN{Dongyue Xue, Eylem Ekici}\\
\IEEEauthorblockA{Department of Electrical and Computer Engineering\\
Ohio State University, USA\\
Email: \{xued, ekici\}@ece.osu.edu}}

\maketitle

\begin{abstract}
In this paper, we propose a cross-layer scheduling algorithm that
achieves a throughput ``$\epsilon$-close'' to the optimal throughput
in multi-hop wireless networks with a tradeoff of
$O(\frac{1}{\epsilon})$ in delay guarantees. The algorithm aims to
solve a joint congestion control, routing, and scheduling problem in
a multi-hop wireless network while satisfying per-flow average
end-to-end delay guarantees and minimum data rate requirements. This
problem has been solved for both backlogged as well as arbitrary
arrival rate systems. Moreover, we discuss the design of a class of
low-complexity suboptimal algorithms, effects of delayed
feedback on the optimal algorithm, and extensions of the
proposed algorithm to different interference models with arbitrary
link capacities.
\end{abstract}

\section{Introduction}
Cross-layer design of congestion control, routing and scheduling
algorithms with Quality of Service (QoS) guarantees is one of the
most challenging topics in wireless networking. The back-pressure
algorithm first proposed in \cite{Lyap0} and its extensions have
been widely employed in developing throughput optimal dynamic
resource allocation and scheduling algorithms for wireless systems.
Back-pressure-based scheduling algorithms have also been employed in
wireless networks with time-varying channels
\cite{Lyap4}\cite{Lyap10}\cite{Lyap11}. Congestion controllers at
the transport layer have assisted the cross-layer design of
scheduling algorithms in \cite{Lyap1}\cite{Lyap2}\cite{Lyap12}, so
that the admitted arrival rate is guaranteed to lie within the
network capacity region. Low-complexity distributed algorithms have
been proposed in \cite{Lyap8}\cite{Lyap9}\cite{dis1}\cite{dis2}.
Algorithms adapted to clustered networks have been proposed in
\cite{Lyap3} to reduce the number of queues maintained in the
network. However, delay-related investigations are not included in
these works.

In this paper, we propose a cross-layer algorithm to achieve
\emph{guaranteed throughput} while satisfying network QoS
requirements. Specifically, we construct two virtual queues, i.e., a
\textit{virtual queue at transport layer} and a \textit{virtual
delay queue}, to \emph{guarantee average end-to-end delay bounds}.
Moreover, we construct a \textit{virtual service queue} to
\emph{guarantee the minimum data rate required by individual network
flows}. Our cross-layer design includes a congestion controller for
the input rate to the virtual queue at transport layer, as well as a
joint policy for packet admission, routing, and resource scheduling.
We show that our algorithm can achieve a throughput arbitrarily
close to the optimal. In addition, the algorithm exhibits a tradeoff
of $O(\frac{1}{\epsilon})$ in the delay bound, where $\epsilon$
denotes the distance from the optimal throughput.

Our main algorithm is further extended: $(1)$ to a set of
low-complexity suboptimal algorithms; $(2)$ from a model with
constantly-backlogged sources to a model with sources of arbitrary
input rates at transport layer; $(3)$ to an algorithm employing
delayed queue information; and $(4)$ from a node-exclusive model with
constant link capacities to a model with arbitrary link capacities
and interference models over fading channels.

The rest of the paper is organized as follows: Section II discusses
the related work. In Section III, the network model is presented,
followed by corresponding approaches for the considered multi-hop
wireless networks. In Section IV, the optimal cross-layer control
and scheduling algorithm is described, and its performance analyzed.
In Section V, we provide a class of feasible suboptimal algorithms,
consider sources with arbitrary arrival rates at transport layer,
employ delayed queue information in the scheduling algorithm, and
extend the model to arbitrary link capacities and interference
models over fading channels. We present numerical results in Section
VI. Finally, we conclude our work in Section VII.

\section{Related Work}
Delay issues in single-hop wireless networks have been addressed in
\cite{delay5}-\cite{plus}. Especially, the scheduling algorithm in
\cite{delay9} provides a throughput-utility that is inversely
proportional to the delay guarantee. Authors of \cite{delay12} have
obtained delay bounds for two classes of scheduling policies. A
random access algorithm is proposed in \cite{csma5} for lattice and
torus interference graphs, which is shown to achieve order-optimal
delay in a distributed manner with optimal throughput. But these
works are not readily extendable to multi-hop wireless networks,
where additional arrivals from neighboring nodes and routing must be
considered. Delay analysis for multi-hop networks with fixed-routing
is provided in \cite{delay1}. Delay-related scheduling in multi-hop
wireless networks have been proposed in
\cite{delay2}\cite{delay3}\cite{delay4}\cite{delay10}\cite{delay11}.
However, none of the above-mentioned works provide explicit
end-to-end delay guarantees.

There are several works aiming to address end-to-end delay or buffer
occupancy guarantees in multi-hop wireless networks. Worst-case
delay is guaranteed in \cite{delay13} with a packet dropping
mechanism. However, dropped packets are not compensated or
retransmitted with the algorithm of \cite{delay13}, which may lead
to restrictions in its practical implementations. A low-complexity
cross-layer fixed-routing algorithm is developed in \cite{csma4} to
guarantee order-optimal average end-to-end delay, but only for half
of the capacity region. A scheduling algorithm for finite-buffer
multi-hop wireless networks with fixed routing is proposed in
\cite{Lyap13} and is extended to adaptive-routing with congestion
controller in \cite{Lyap14}. Specifically, the algorithm in
\cite{Lyap14} guarantees $O(\frac{1}{\epsilon})$-scaling in buffer
size with a $\epsilon$-loss in throughput-utility, but this is
achieved at the expense of the buffer occupancy of the source nodes,
where \emph{an infinite buffer size} in the network layer is assumed
in each source node. This leads to large average end-to-end delay
since the network stability is achieved based on queue backlogs at
these source nodes.

Compared to the above works, the algorithm presented in this paper
develops and incorporates novel virtual queue structures. Different
from traditional back-pressure-based algorithms, where the network
stability is achieved at the expense of large packet queue backlogs,
in our algorithm, ``the burden'' of actual packet queue backlogs is
shared by our proposed virtual queues, in an attempt to guarantee
specific delay performances. Specifically, we design a congestion
controller for \emph{a virtual input rate} and assign weights in the
scheduling policy as a product of actual packet queue backlog and
the weighted backlog of a designed virtual queue, which will be
introduced in detail in Section IV. As such, the network
stabilization is achieved with the help of virtual queue structures
that do not contribute to delay in the network. Since \textit{all
packet queues} in the network, including those in source nodes, have
finite sizes, all average end-to-end delays are bounded independent
of length or multiplicity of paths.

\section{Network Model}
\subsection{Network Elements}
We consider a time-slotted multi-hop wireless network consisting of
$N$ nodes and $K$ flows. Denote by $(m,n)\in \mathcal {L}$ a link
from node $m$ to node $n$, where $\mathcal{L}$ is the set of
directed links in the network. Denoting the set of flows by
$\mathcal{F}$ and the set of nodes by $\mathcal{N}$, we formulate
the network topology $G=(\mathcal{N},\mathcal{L})$. Note that we
consider adaptive routing scenario, i.e., the routes of each flow
are not determined \emph{a priori}, which is more general than
fixed-routing scenario. In addition, we denote the source node and
the destination node of a flow $c\in\mathcal{F}$ as $b(c)$ and
$d(c)$, respectively.

We assume that the source node for flow $c$ is always backlogged at
the transport layer. Let the scheduling parameter $\mu_{mn}^{c}(t)$
denote the link rate assignment of flow $c$ for link $(m,n)$ at time
slot $t$ according to scheduling decisions and let
$\mu_{s(c)b(c)}^{c}(t)$ denote the admitted rate of flow $c$ from
the transport layer of flow to the source node, where $s(c)$ denotes
the source at the transport layer of flow $c$. It is clear that in
any time slot $t$, $\mu_{s(c)n}^{c}(t)=0$ $\forall n\neq b(c)$. For
simplicity of analysis, we assume only one packet can be transmitted
over a link in one slot, so $(\mu_{mn}^c(t))$ takes values in
$\{0,1\}$ $\forall (m,n)\in \mathcal{L}$. We also assume that
$\mu_{s(c)b(c)}^c(t)$ is bounded above by a constant $\mu_M\geq1$:
\begin{equation}\label{e:3}
0\leq\mu_{s(c)b(c)}^{c}(t)\leq \mu_M, \mbox{ }\forall
c\in\mathcal{F},\forall t,
\end{equation}
i.e., a source node can receive at most $\mu_M$ packets from the
transport layer in any time slot. To simplify the analysis, we
prevent looping back to the source, i.e., we impose the following
constraints
\begin{equation}\label{e:10}
\sum_{m\in\mathcal{N}}(\mu_{mb(c)}^c(t))=0 \mbox{ }\forall
c\in\mathcal{F},\forall t.
\end{equation}
We employ the node-exclusive model in our analysis, i.e., each node
can communicate with at most one other node in a time slot. Note
that our model is extended to arbitrary interference models with
arbitrary link capacities and fading channels in Section V.D.

We now specify the QoS requirements associated with each flow. The
network imposes an \emph{average end-to-end delay threshold}
$\rho_c$ for each flow $c$. The end-to-end delay period of a packet
starts when the packet is admitted to the source node from the
transport layer and ends when it reaches its destination. Note that
the delay threshold is a time-averaged upper-bound, not a
deterministic one. In addition, each flow $c$ requires a minimum
data rate of $a_c$ packets per time slot.

\subsection{Network Constraints and Approaches}
For convenience of analysis, we define $\mathcal{L}^c\triangleq
\mathcal{L}\cup\{(s(c),b(c))\}$, where the pair $(s(c), b(c))$ can
be considered as a virtual link from transport layer to the source
node. We now model queue dynamics and network constraints in the
multi-hop network. Let $U_n^c(t)$ be the backlog of the total amount
of flow $c$ packets waiting for transmission at node $n$. For a flow
$c$, if $n=d(c)$ then $U_n^c(t)=0$ $\forall t$; Otherwise, the queue
dynamics is as follows:
\begin{eqnarray}\label{eq:1}
\begin{aligned}
U_n^c(t+1) \leq &[U_n^c(t)-\sum_{i: (n,i)\in \mathcal
{L}}\mu_{ni}^{c}(t)]^+&\\
+&\sum_{j: (j,n)\in \mathcal{L}^c}\mu_{jn}^{c}(t), \mbox{ if }n\in
\mathcal{N}\backslash d(c),&\\
\end{aligned}
\end{eqnarray}
where the operator $[x]^+$ is defined as $[x]^+=\max\{x,0\}$. Note
that in (\ref{eq:1}), we ensure that the actual number of packets
transmitted for flow $c$ from node $n$ does not exceed its queue
backlog, since a feasible scheduling algorithm may not depend on the
information on queue backlogs. The terms $\sum_{i: (n,i)\in \mathcal
{L}}\mu_{ni}^{c}(t)$ and $\sum_{j: (j,n)\in \mathcal
{L}^c}\mu_{jn}^{c}(t)$ represent, respectively, the scheduled
departure rate from node $n$ and the scheduled arrival rate into
node $n$ by the scheduling algorithm with respect to flow $c$. Note
that (\ref{eq:1}) is an inequality since the arrival rates from
neighbor nodes may be less than $\sum_{j}\mu_{jn}^{c}(t)$ if some
neighbor node does not have sufficient number of packets to
transmit. Since we employ the node-exclusive model, we have
\begin{equation}\label{e:4}
0\leq \sum_{c\in\mathcal{F}}[\sum_{i: (n,i)\in \mathcal
{L}}\mu_{ni}^{c}(t)+\sum_{j: (j,n)\in \mathcal
{L}}\mu_{jn}^{c}(t)]\leq 1 \mbox{, }\forall n\in\mathcal{N}.
\end{equation}
From (\ref{e:3})(\ref{e:10}), we also have
\begin{equation}\label{e:5}
\sum_{j:(j,n)\in\mathcal{L}^c}\mu_{jn}^{c}(t)\leq \mu_M \mbox{, if }
n=b(c),
\end{equation}
if it is ensured that no packets will be looped back to the source.

Now we construct three kinds of virtual queues, namely, virtual
queue $U_{s(c)}^c(t)$ at transport layer, virtual service queue
$Z_c(t)$ at sources, and virtual delay queue $X_c(t)$, to later
assist the development of our algorithm:
\\
$(1)$ For each flow $c$ at transport layer, we construct a virtual
queue $U_{s(c)}^c(t)$ which will be employed in the algorithm
proposed in the next section. We denote the virtual input rate to
the queue as $R_c(t)$ at the end of time slot $t$ and we upper-bound
$R_c(t)$ by $\mu_M$. Let $r_c$ denote the time-average of $R_c(t)$.
We update the virtual queue as follows:
\begin{eqnarray}\label{eq:4}
U_{s(c)}^c(t+1) = [U_{s(c)}^c(t)-\mu_{s(c)b(c)}^{c}(t)]^+ +R_c(t),
\end{eqnarray}
where the initial $U_{s(c)}^c(0)=0$. Considering the admitted rate
$\mu_{s(c)b(c)}^{c}(t)$ as the service rate, if the virtual queue
$U_{s(c)}^c(t)$ is stable, then the time-average admitted rate
$\mu_c$ of flow $c$ satisfies:
\begin{equation}\label{e:0}
\mu_c\triangleq\lim_{t\rightarrow
\infty}\frac{1}{t}\sum_{\tau=0}^{t-1}\mu_{s(c)b(c)}^{c}(\tau)\geq
r_c\triangleq\lim_{t\rightarrow
\infty}\frac{1}{t}\sum_{\tau=0}^{t-1}R_c(\tau).
\end{equation}
\\
$(2)$ To satisfy the minimum data rate constraints, we construct a
virtual queue $Z_c(t)$ associated with flow $c$ as follows:
\begin{equation}\label{e:1}
Z_c(t+1)=[Z_c(t)-R_c(t)]^+ +a_c,
\end{equation}
where the initial $Z_c(0)=0$. Considering $a_c$ as the arrival rate
and $R_c(t)$ as the service rate, if queue $Z_c(t)$ is stable, we
have: $r_c\geq a_c$. Additionally, if $U_{s(c)}^c(t)$ is stable,
then according to (\ref{e:0}), the minimum data rate for flow $c$ is
achieved.
\\
$(3)$ To satisfy the end-to-end delay constraints, we construct a
virtual delay queue $X_c(t)$ for any given flow $c$ as follows:
\begin{equation}\label{eq:2}
X_c(t+1)=[X_c(t)-\rho_cR_c(t)]^++\sum_{n\in \mathcal{N}}U_{n}^c(t)
\end{equation}
where the initial $X_c(0)=0$. 
Considering the packets kept in the network in time slot $t$, i.e.,
$\sum_{n\in \mathcal{N}}U_{n}^c(t)$, as the arrival rate and
$\rho_cR_c(t)$ as the service rate, and according to queueing
theory, if queue $X_c(t)$ is stable, we have
\begin{displaymath}
\lim_{t\rightarrow \infty} \frac{1}{t}\sum_{\tau=0}^{t-1}\sum_{n\in
\mathcal{N}}U_n^c(\tau)\leq \rho_c\lim_{t\rightarrow
\infty}\frac{1}{t}\sum_{\tau=0}^{t-1}R_c(\tau)=\rho_cr_c.
\end{displaymath}
Furthermore, if $U_{s(c)}^c(t)$ is stable, then according to
(\ref{e:0}), we have:
\begin{equation}\label{eq:3}
\frac{1}{\mu_c}\lim_{t\rightarrow
\infty}\frac{1}{t}\sum_{\tau=0}^{t-1}\sum_{n\in
\mathcal{N}}U_n^c(\tau)\leq \rho_c.
\end{equation}
In addition, by Little's Theorem, (\ref{eq:3}) ensures that the
average end-to-end delay of flow $c$ is less than or equal to the
threshold $\rho_c$ with probability (w.p.) $1$. 

From the above description, we know that the network is
\emph{stable} (i.e., each queue at all nodes is stable) and the
average end-to-end delay constraint and minimum data rate
requirement are achieved if queues $U_n^c(t)$ and the three virtual
queues are stable for any node and flow, i.e.,

\begin{displaymath}
\limsup_{t\rightarrow \infty}
\frac{1}{t}\sum_{\tau=0}^{t-1}\mathbb{E}\{X_c(\tau)\} <\infty,\quad
\forall c;
\end{displaymath}
\begin{displaymath}
\limsup_{t\rightarrow \infty}\frac{1}{t}
\sum_{\tau=0}^{t-1}\mathbb{E}\{U_n^c(\tau)\} <\infty,\quad \forall
n\in\mathcal{N}\cup\{s(c):c\in\mathcal{F}\};
\end{displaymath}
\begin{displaymath}
\limsup_{t\rightarrow \infty}
\frac{1}{t}\sum_{\tau=0}^{t-1}\mathbb{E}\{Z_c(\tau)\} <\infty,\quad
\forall c.
\end{displaymath}

Now we define the capacity region of the considered multi-hop
network. An arrival rate vector $(z_c)$ is called \emph{admissible}
if there exists some scheduling algorithm (without congestion
control) under which the node queue backlogs (not including virtual
queues) are stable. We denote $\Lambda$ to be the capacity region
consisting of all admissible $(z_c)$, i.e., $\Lambda$ consists of
all feasible rates stabilizable by some scheduling algorithm
\emph{without} considering QoS requirements (i.e., delay constraints
and minimum data rate constraints). To assist the analysis in the
following sections, we let $(r_{\epsilon,c}^*)$ denote the solutions
to the following optimization problem:
\begin{displaymath}
\max_{(r_c):(r_c+\epsilon)\in\Lambda}\sum_{c\in\mathcal{F}}r_c
\end{displaymath}
\begin{displaymath}
\mbox{s.t. } r_c\geq a_c, \mbox{ } \forall c\in\mathcal{F}.
\end{displaymath}
where $\epsilon$ is a positive number which can be chosen
arbitrarily small. For simplicity of analysis, we assume that
$(a_c)$ is in the interior of $\Lambda$ 
and without loss of generality, we assume that there exists
$\epsilon'>0$ such that $r_{\epsilon,c}^*\geq a_c+\epsilon'$
$\forall c\in\mathcal{F}$. According to \cite{Lyap6}, we have
\begin{displaymath}
\lim_{\epsilon\rightarrow0}\sum_{c\in\mathcal{F}}r_{\epsilon,c}^*=\sum_{c\in\mathcal{F}}r_c^*,
\end{displaymath}
where $(r_c^*)$ is the solution to the following optimization:
\begin{displaymath}
\begin{aligned}
&\max_{(r_c):(r_c)\in\Lambda}\sum_{c\in\mathcal{F}}r_c&\\
&\mbox{s.t. } r_c\geq a_c, \mbox{ } \forall c\in\mathcal{F}.&
\end{aligned}
\end{displaymath}

\section{Control Scheduling Algorithm for Multi-Hop Wireless Networks}
Now we propose a control and scheduling algorithm
\emph{\textbf{ALG}} for the introduced multi-hop model so that
\emph{\textbf{ALG}} stabilizes the network and satisfies the delay
constraint and minimum data rate constraint. Given $\epsilon$, the
proposed \emph{\textbf{ALG}} can achieve a throughput arbitrarily
close to $\sum_{c\in\mathcal{F}}r_{\epsilon,c}^*$, under certain
conditions related to delay constraints which will be later given in
Theorem 1.

The optimal algorithm \emph{\textbf{ALG}} consists of two parts: a
congestion controller of $R_c(t)$, and a joint packet admission,
routing and scheduling policy. We propose and analyze the algorithm
in the following subsections.

\subsection{Algorithm Description and Analysis}
Let $q_M\geq\mu_M$ be a control parameter for queue length. We first
propose a congestion controller for the input rate of virtual queues
at transport layer:

\textbf{1) Congestion Controller of $R_c(t)$}:
\begin{equation}\label{eq:25}
\min_{0\leq R_c(t)\leq\mu_M}
R_c(t)(\frac{(q_M-\mu_{M})U_{s(c)}^c(t)}{q_M}-X_c(t)\rho_c-Z_c(t)-V)
\end{equation}
where $V>0$ is a control parameter. Specifically, when
$\frac{q_M-\mu_{M}}{q_M}U_{s(c)}^c(t)-X_c(t)\rho_c-Z_c(t)-V>0$,
$R_c(t)$ is set to zero; Otherwise, $R_c(t)=\mu_M$.

After performing the congestion control, we perform the following
joint policy for packet admission, routing and scheduling
(abbreviated as \emph{scheduling policy}):

\textbf{2) Scheduling Policy}: In each time slot, with the
constraints of the underlying interference model as described in
Section III including (\ref{e:3})(\ref{e:10})(\ref{e:4}), the
network solves the following optimization problem:

\begin{equation}\label{eq:11}
\max_{(\mu_{mn}^c(t))}\sum_{m,n}\mu_{mn}^{c_{mn}^*(t)}(t)w_{mn}(t)
\end{equation}
\begin{displaymath}
\mbox{s.t. } \quad \mu_{mn}^{c}(t)=0 \quad\mbox{ }\forall c\neq
c_{mn}^*(t),\mbox{ } \forall (m,n)\in\mathcal{L}^c,
\end{displaymath}
\begin{displaymath}
\quad\mu_{mn}^{c}(t)=0 \quad\mbox{if } n=s(c) \mbox{, }\forall
c\in\mathcal{F} ,\qquad
\end{displaymath}
where $c_{mn}^*(t)$ and $w_{mn}(t)$ are defined as follows:
\begin{displaymath}
c_{mn}^*(t)=\mbox{arg}\max_{c\in\mathcal{F}}w_{mn}^c(t),
\end{displaymath}
\begin{displaymath}
w_{mn}(t)=[\max_{c\in\mathcal{F}}w_{mn}^c(t)]^+,
\end{displaymath}
with weight assignment as follows
\begin{eqnarray}\label{e:6}
w_{mn}^c(t)=\left\{\begin{aligned} &
\frac{U_{s(c)}^c(t)}{q_M}[U_m^c(t)-U_n^c(t)],\mbox{ if }
(m,n)\in \mathcal{L}, &\\
& \frac{U_{s(c)}^c(t)}{q_M}[q_M-\mu_M-U_{b(c)}^c(t)],&\\
&\qquad\qquad\qquad\mbox{ if }
(m,n)=(s(c),b(c)), &\\
&0, \qquad\qquad\quad\mbox{ otherwise. }&\\
\end{aligned} \right.
\end{eqnarray}
In addition, when $w_{mn}(t)=0$, without loss of optimality, we set
$\mu_{mn}^{c}(t)=0$ $\forall c\in\mathcal{F}$ to maximize
(\ref{eq:11}).

Note that $\mathcal{L}\cup\{(s(c),b(c)):c\in\mathcal{F}\}$ forms the
$(m,n)$ pairs in $(\mu_{mn}^c(t))$ over which the optimization
(\ref{eq:11}) is performed. Thus, the optimization is a typical
Maximum Weight Matching (MWM) problem. We first decouple flow
scheduling from the MWM. Specifically, for each pair $(m,n)$, the
flow $c_{mn}^*(t)$ is fixed as the candidate for transmission. We
then assign the weight as $w_{mn}(t)$.  Note also that although
similar product form of the weight assignment (\ref{e:6}) have been
utilized in \cite{Lyap13}\cite{Lyap14}, no virtual queues are
involved there. Whereas in \emph{\textbf{ALG}}, we assign weights as
a product of weighted virtual queue backlog
($\frac{U_{s(c)}^c(t)}{q_M}$) and the actual back-pressure, in an
aim to shift the burden of the actual queue backlog to the virtual
backlog.

To analyze the performance of the algorithm, we first introduce the
following proposition.

\begin{proposition}
Employing \emph{\textbf{ALG}}, each queue backlog in the network has
a deterministic worst-case bound:
\begin{equation}\label{eq:26}
U_n^c(t)\leq q_M, \quad \forall t,\forall n\in\mathcal{N},\forall
c\in\mathcal{F}.
\end{equation}
\end{proposition}
\proof{We use mathematical induction on time slot in the proof. When
$t=0$, $U_n^c(0)=0\leq q_M$ $\forall n,c$. In the induction
hypothesis, we suppose in time slot $t$ we have $U_n^c(t)\leq q_M$
$\forall n,c$. In the induction step, for any given
$n\in\mathcal{N}$ and $c\in \mathcal{F}$, we consider two cases as
follows:
\\
$(1)$ We first consider the case when $n=b(c)$, i.e., when $n$ is
the source node of flow $c$. Since $U_n^c(t)\leq q_M$ from the
induction hypothesis, we further consider two subcases:
\begin{itemize}
\begin{item}
In the first subcase, $U_{b(c)}^c(t)\leq q_M-\mu_M$. Then according
to the queue dynamics (\ref{eq:1}) and the inequality (\ref{e:5}),
$U_{b(c)}^c(t+1)\leq U_{b(c)}^c(t)+\mu_M\leq q_M$;
\end{item}
\begin{item}
In the second subcase, $q_M-\mu_M<U_{b(c)}^c(t)\leq q_M $. According
to the weight assignment (\ref{e:6}), we have $w_{s(c)b(c)}^c(t)<0$
which leads to $\mu_{s(c)b(c)}^c(t)=0$. Hence, $U_{b(c)}^c(t+1)\leq
U_{b(c)}^c(t)\leq q_M$ by (\ref{e:10})(\ref{eq:1}).
\end{item}
\end{itemize}
$(2)$ In the second case, $n\neq b(c)$, i.e., $n$ is not the source
node of flow $c$. Similar to the first case, we further consider the
following two subcases:
\begin{itemize}
\begin{item}
In the first subcase, $U_n^c(t)< q_M$. Then, since we employ
node-exclusive model, $U_n^c(t+1)\leq U_n^c(t)+1\leq q_M$ by
(\ref{eq:1})(\ref{e:4}).
\end{item}
\begin{item}
In the second subcase, $U_n^c(t)= q_M$. According to the weight
assignment (\ref{e:6}) we have $w_{mn}^c(t)\leq 0$ $\forall m:
(m,n)\in\mathcal{L}$. Now, for any given node $m:
(m,n)\in\mathcal{L}$, we have:
\\
(i) If $c\neq c_{mn}^*(t)$, then by (\ref{eq:11}),
$\mu_{mn}^{c}(t)=0$;
\\
(ii) Otherwise, $c=c_{mn}^*(t)$, which induces
$w_{mn}(t)=[w_{mn}^c(t)]^+=0$ and by the scheduling policy,
$\mu_{mn}^{c}(t)=0$.
\\
Hence $\mu_{mn}^c(t)=0$ $\forall m: (m,n)\in\mathcal{L}$, and
$U_n^c(t+1)\leq U_n^c(t)=q_M$ by the queue dynamics (\ref{eq:1}).
\end{item}
\end{itemize}
The above analysis holds for any given $n\in\mathcal{N}$ and
$c\in\mathcal{F}$. Therefore the induction step holds, i.e.,
$U_n^c(t+1)\leq q_M$ $\forall n,c$, which completes the proof.}
\endproof

Now we present our main results in Theorem 1. \theorem{Given that
\begin{equation}\label{eq:7}
q_M> \frac{2N-1+\mu_M^2}{2\epsilon}+\mu_M \mbox{ and } \rho_c>
\frac{Nq_M}{r_{\epsilon,c}^*} \mbox{ } \forall c\in\mathcal{F},
\end{equation}
\emph{\textbf{ALG}} can achieve a throughput
\begin{equation}\label{eq:16}
\liminf_{t\rightarrow\infty}\frac{1}{t}\sum_{\tau=0}^{t-1}\sum_{c\in\mathcal{F}}\mathbb{E}\{R_c(\tau)\}\geq\sum_{c\in\mathcal{F}}r_{\epsilon,c}^*-\frac{B}{V},
\end{equation}
where
$B\triangleq\frac{1}{2}NKq_M\mu_M+K\frac{q_M-\mu_M}{q_M}\mu_M^2+\frac{1}{2}\mu_M^2\sum_{c\in\mathcal{F}}\rho_c^2+\frac{1}{2}KN^2q_M^2+\frac{1}{2}K\mu_M^2+\frac{1}{2}K\sum_{c\in\mathcal{F}}a_c^2$.

In addition, \emph{\textbf{ALG}} ensures that the virtual queues
have a time-averaged bound:
\begin{equation}\label{eq:14}
\limsup_{t\rightarrow
\infty}\frac{1}{t}\sum_{\tau=0}^{t-1}\sum_{c\in\mathcal{F}}\mathbb{E}\{U_{s(c)}^c(\tau)+X_c(\tau)+Z_c(\tau)\}\leq\frac{B'}{\delta},
\end{equation}
where $B'\triangleq B+VB_R$, with $B_R$ and $\delta$ constant
positive numbers given in the next subsection. }

\emph{Remark 1 (Network Stability)}: The inequalities (\ref{eq:26})
from Proposition 1 and (\ref{eq:14}) from Theorem 1 indicate that
\emph{\textbf{ALG}} stabilizes the actual and virtual queues. As an
immediate result, \emph{\textbf{ALG}} stabilizes the network and
satisfies the average end-to-end delay constraint and the minimum
data rate requirement. In addition, Proposition 1 states that the
actual queues are \emph{deterministically} bounded by $q_M$, which
ensures finite buffer sizes for all queues in the network, including
those in source nodes.

\emph{Remark 2 (Optimal Utility and Delay Analysis): } Since
$(U_{s(c)}^c(t))$ are stable, the inequality (\ref{eq:16}) gives a
lower-bound on the throughput that \emph{\textbf{ALG}} can achieve.
Given some $ \epsilon>0$, since $B$ is independent of $V$,
(\ref{eq:16}) also ensures that \emph{\textbf{ALG}} can achieve a
throughput arbitrarily close to
$\sum_{c\in\mathcal{F}}r_{\epsilon,c}^*$. When $\epsilon$ tends to
$0$, \emph{\textbf{ALG}} can achieve a throughput arbitrarily close
to the optimal value $\sum_{c\in\mathcal{F}}r_{c}^*$ with the
tradeoff in queue backlog upper-bound $q_M$ and the delay
constraints $(\rho_c)$, both of which are lower-bounded by the
reciprocal terms of $\epsilon$ as shown in (\ref{eq:7}) in Theorem
1. \textit{In other words, the average end-to-end delay bound is of
order $O(\frac{1}{\epsilon})$}. We note that in \emph{\textbf{ALG}},
the control parameter $V$, which is typically chosen to be large,
does not affect the actual queue backlog upper-bound or the average
end-to-end delay bound, but only affects the upper-bound of the
virtual queue backlogs (shown in (\ref{eq:14})). In comparison, in
the algorithm proposed in \cite{Lyap14}, the authors show that the
internal buffer size is deterministically bounded with order
$O(\frac{1}{\epsilon})$, but \emph{at the expense of} the buffer
occupancy at source nodes which is of order $O(V)$, where $V$ has to
be large enough for their algorithm to approach
$\sum_{c\in\mathcal{F}}r_{\epsilon,c}^*$. This design assumes an
\emph{infinite buffer size} at source nodes and typically results in
congestion at the source nodes as shown in the simulation results in
\cite{Lyap14}, which further induces an unguaranteed and large
average end-to-end delay. Moreover, one can expect that there are no
buffer-size guarantees for single-hop flows by employing the
algorithm in \cite{Lyap14}. In contrast, in our proposed
\emph{\textbf{ALG}}, we shift ``the burden of $V$'' from actual
queues to virtual queues and ensure that the average end-to-end
delay constraints are satisfied with finite buffer sizes for all
actual \textit{packet} queues.

\emph{Remark 3 (Implementation Issues)}: To update the virtual queue
$X_c(t)$ and perform the $R_c(t)$ congestion controller at the
transport layer, the queue backlog information of flow $c$ is
crucial. This information can be collected back to the source node
by piggy-backing it on ACK from each node. In order to account for
such delay of queue backlog information, the $R_c(t)$ congestion
controller (\ref{eq:25}) of the algorithm can employ delayed queue
backlog of $X_c(t)$. Similarly, delayed queue backlog information of
$U_{s(c)}^c(t)$ can be employed at the weight assignment (\ref{e:6})
of the scheduling policy. The modified algorithm and its validity
are further discussed in Section V.C. By employing delayed queue
backlog information, we can extend the algorithm to distributed
implementation in much the same way as in \cite{Lyap8}\cite{dis2} to
achieve \emph{a fraction} of the optimal throughput. In order to
achieve a throughput arbitrarily close to the optimal value with
distributed implementation, we can employ random access techniques
\cite{csma1}\cite{csma2} in the scheduling policy with fugacities
\cite{csma3} chosen as
exp$\{\frac{\alpha\bar{U}_{s(c)}^c(t)[U_m^c(t)-U_n^c(t)]^+}{q_M}\}$
for each link $(m,n)\in\mathcal{L}$, where $\bar{U}_{s(c)}^c(t)$ is
a local estimate (e.g., delayed information) of $U_{s(c)}^c(t)$ and
$\alpha$ a positive weight. It can be shown that the distributed
algorithm can still achieve an average end-to-end delay of order
$O(\frac{1}{\epsilon})$ with the time-scale separation assumption
\cite{csma5}\cite{csma0}\cite{csma1}.\footnote{Note that the random
access works cited above either do not provide delay guarantees or
are not readily extended to multi-hop settings.} A variation of such
distributed implementation in single-hop networks can be found in
our recent work \cite{report}.


We prove Theorem 1 in the following subsection.

\subsection{Proof of Theorem 1}
Before we proceed, we present the following lemmas which will assist
us in proving Theorem 1.

\lemma{ For nonnegative numbers $A_1,A_2,A_3,Q \in \mathbb{R}$ such
that $Q\leq [A_1-A_2]^++A_3$, we have $Q^2\leq
A_1^2+A_2^2+A_3^2-2A_1(A_2-A_3).$ }

The proof of Lemma 1 is trivial and omitted. We will later use Lemma
1 to simplify virtual queue dynamics.

\lemma{For any feasible rate vector $(\theta_c)\in\Lambda$ with
$\theta_c\geq a_c$ $\forall c\in\mathcal{F}$, there exists a
stationary randomized algorithm STAT that stabilizes the network
with input rate vector $(\mu_{s(c)b(c)}^{STAT}(t))$ and scheduling
parameters $(\mu_{mn}^{c,STAT}(t))$ independent of queue backlogs,
such that the expected admitted rates are:
\begin{displaymath}
\mathbb{E}\{\mu_{s(c)b(c)}^{c,STAT}(t)\}=\theta_c, \forall t,\forall
c\in\mathcal{F}.
\end{displaymath}
In addition, $\forall t$, $\forall n\in\mathcal{N},\forall c$, the
flow constraint is satisfied:
\begin{displaymath}
\mathbb{E}\{\sum_{i:(n,i)\in\mathcal{L}}\mu_{ni}^{c,STAT}(t)-\sum_{j:(j,n)\in\mathcal{L}^c}\mu_{jn}^{c,STAT}(t)\}=0.
\end{displaymath}}

Note that it is not necessary for the randomized algorithm STAT to
satisfy the average end-to-end delay constraints. Similar
formulations of STAT and their proofs have been given in
\cite{Lyap1} and \cite{Lyap2}, so we omit the proof of Lemma 2 for
brevity.

\emph{Remark 4}: According to the STAT algorithm in Lemma 2, we
assign the input rates of the virtual queues at transport layer as
$R_c^{STAT}(t)=\mu_{s(c)b(c)}^{c,STAT}(t)$. Thus, we also have
$\mathbb{E}\{R_c^{STAT}(t)\}=\theta_c$. According to the update
equation (\ref{eq:4}), it is easy to show that the virtual queues
under STAT are bounded above by $\mu_M$ and the time-average of
$R_c^{STAT}(t)$ satisfies: $r_c^{STAT}=\theta_c$. Note that
$(\theta_c)$ can take values as $(r_{\epsilon,c}^*)$ or
$(r_{\epsilon,c}^*+\epsilon)$ or
$(r_{\epsilon,c}^*-\frac{1}{2}\epsilon')$, where we recall
$(r_{\epsilon,c}^*+\epsilon)\in\Lambda$ and $r_{\epsilon,c}^*\geq
a_c+\epsilon'$ $\forall c\in\mathcal{F}$.

To prove Theorem 1, we first let
$\textbf{Q}(t)=((U_n^c(t)),(U_{s(c)}^c(t)),(X_c(t)),(Z_c(t)))$ and
define the Lyapunov function $L(\textbf{Q}(t))$ as follows:
\begin{eqnarray}\label{eq:18}
\begin{aligned}
L(\textbf{Q}(t))&=\frac{1}{2}\{\sum_{c\in\mathcal{F}}\frac{q_M-\mu_M}{q_M}{U_{s(c)}^c(t)}^2+\sum_{c\in\mathcal{F}}X_c(t)^2&\\
&+\sum_{c\in\mathcal{F}}Z_c(t)^2
+\sum_{c\in\mathcal{F}}\sum_{n\in\mathcal{N}}\frac{1}{q_M}U_n^c(t)^2U_{s(c)}^c(t)\}.&\\
\end{aligned}
\end{eqnarray}
It is obvious that $L(\textbf{Q}(0))=0$. We denote the Lyapunov
drift by
\begin{equation}\label{eq:12}
\Delta(t)=\mathbb{E}\{L( \textbf{Q}(t+1)) -L(\textbf{Q}(t))
|\textbf{Q}(t)\}.
\end{equation}
From the queue dynamics
(\ref{eq:1})(\ref{eq:4}), we have:
\begin{eqnarray}\label{e:16}
\begin{aligned}
&\sum_{c\in\mathcal{F}}\sum_{n\in\mathcal{N}}\frac{1}{q_M}U_n^c(t+1)^2U_{s(c)}^c(t+1)&\\
\leq&\sum_{c\in\mathcal{F}}\frac{1}{q_M}(R_c(t)+U_{s(c)}^c(t))\sum_{n\in\mathcal{N}}U_n^c(t+1)^2 &\\
\leq&\mu_Mq_MNK+\sum_{c\in\mathcal{F}}\frac{1}{q_M}U_{s(c)}^c(t)\sum_{n\in\mathcal{N}}\{U_n^c(t)^2&\\
&\quad+(\sum_{i:(n,i)\in\mathcal{L}}
\mu_{ni}^c(t))^2+(\sum_{j:(j,n)\in\mathcal{L}^c}\mu_{jn}^c(t))^2&\\
&\qquad-2U_n^c(t)(\sum_{i}
\mu_{ni}^c(t)-\sum_{j}\mu_{jn}^c(t))\},&\\
\end{aligned}
\end{eqnarray}
where we recall that $R_c(t)\leq\mu_M$ and we employ Lemma 1 to
deduce the second inequality.

From (\ref{e:16}), we have
\begin{eqnarray}\label{e:11}
\begin{aligned}
&
\frac{1}{2}(\sum_{c\in\mathcal{F}}\sum_{n\in\mathcal{N}}\frac{1}{q_M}(U_n^c(t+1)^2U_{s(c)}^c(t+1)&\\
&\qquad\qquad\qquad\quad-U_n^c(t)^2U_{s(c)}^c(t)))&\\
\leq&\frac{1}{2}\sum_{c\in\mathcal{F}}\frac{(2N-1+\mu_M^2)U_{s(c)}^c(t)}{q_M}+\frac{1}{2}NKq_M\mu_M&\\
-&\sum_{c\in\mathcal{F}}\sum_{n\in\mathcal{N}}\frac{U_n^c(t)U_{s(c)}^c(t)}{q_M}&\\
&\qquad(\sum_{j:(n,j)\in\mathcal{L}}\mu_{nj}^c(t)-\sum_{i:(i,n)\in\mathcal{L}^c}\mu_{in}^c(t)),&\\
\end{aligned}
\end{eqnarray}
where we employ the fact deduced from (\ref{e:4})(\ref{e:5}) that
$\sum_{i} \mu_{ni}^c(t)\leq1$ and $\sum_{j}\mu_{jn}^c(t)\leq 1$ when
$n\neq b(c)$ and $\sum_{j}\mu_{jn}^c(t)\leq \mu_M$ when $n=b(c)$.
Note that we use the summation index $i$ and $j$ interchangeably for
convenience of analysis.

From the queue length dynamics (\ref{eq:4}) and by employing Lemma
1, we have:
\begin{eqnarray}\label{eq:6}
\begin{aligned}
&\frac{1}{2}\sum_{c\in\mathcal{F}}\frac{q_M-\mu_M}{q_M}({U_{s(c)}^c(t+1)}^2-{U_{s(c)}^c(t)}^2)&\\
\leq&\frac{1}{2}\sum_{c\in\mathcal{F}}\frac{q_M-\mu_M}{q_M}(\mu_{s(c)b(c)}^c(t)^2+R_c(t)^2&\\
&-2U_{s(c)}^c(t)(\mu_{s(c)b(c)}^c(t)-R_c(t)))       &\\
\leq&K\frac{q_M-\mu_M}{q_M}\mu_M^2&\\
&-\frac{q_M-\mu_M}{q_M}\sum_{c\in\mathcal{F}}U_{s(c)}^c(t)(\mu_{s(c)b(c)}^c(t)-R_c(t)). &\\
\end{aligned}
\end{eqnarray}
From the virtual queue dynamics (\ref{eq:2}), we have:
\begin{eqnarray}\label{e:12}
\begin{aligned}
&\frac{1}{2}\sum_{c\in\mathcal{F}}(X_c(t+1)^2-X_c(t)^2)&\\
\leq&\frac{1}{2}\sum_{c\in\mathcal{F}}(\rho_c^2R_c(t)^2+(\sum_{n\in\mathcal{N}}U_n^c(t))^2&\\
&-2X_c(t)(\rho_cR_c(t)-\sum_{n\in\mathcal{N}}U_n^c(t)))&\\
\leq&\frac{1}{2}\mu_M^2\sum_{c\in\mathcal{F}}\rho_c^2+\frac{1}{2}KN^2q_M^2&\\
&-\sum_{c\in\mathcal{F}}X_c(t)\rho_cR_c(t)+Nq_M\sum_{c\in\mathcal{F}}X_c(t).  &\\
\end{aligned}
\end{eqnarray}
From the virtual queue dynamics (\ref{e:1}), we have:
\begin{eqnarray}\label{e:13}
\begin{aligned}
&\frac{1}{2}\sum_{c\in\mathcal{F}}(Z_c(t+1)^2-Z_c(t)^2)&\\
\leq&\frac{1}{2}\sum_{c\in\mathcal{F}}(R_c(t)^2+a_c^2-2Z_c(t)(R_c(t)-a_c))&\\
\leq&\frac{1}{2}K\mu_M^2+\frac{1}{2}\sum_{c\in\mathcal{F}}a_c^2-\sum_{c\in\mathcal{F}}Z_c(t)R_c(t)+\sum_{c\in\mathcal{F}}a_cZ_c(t).  &\\
\end{aligned}
\end{eqnarray}
Substituting (\ref{e:11})(\ref{eq:6})(\ref{e:12})(\ref{e:13}) into
the Lyapunov drift (\ref{eq:12}) and subtracting
$V\sum_{c}\mathbb{E}\{R_c(t)|\textbf{Q}(t)\}$ from both sides, we
then have:
\begin{eqnarray}\label{e:14}
\begin{aligned}
&\Delta(t)-V\sum_{c\in\mathcal{F}}\mathbb{E}\{R_c(t)|\textbf{Q}(t)\}& \\
\leq&B+\sum_{c\in\mathcal{F}}\mathbb{E}\{R_c(t)(\frac{(q_M-\mu_{M})U_{s(c)}^c(t)}{q_M}&\\
&\qquad\qquad\qquad-X_c(t)\rho_c-Z_c(t)-V)|\textbf{Q}(t)\}&\\
+&Nq_M\sum_{c\in\mathcal{F}}X_c(t)+\sum_{c\in\mathcal{F}}a_cZ_c(t)&\\
+&\frac{1}{2}\sum_{c\in\mathcal{F}}\frac{(2N-1+\mu_M^2)U_{s(c)}^c(t)}{q_M} &\\
-&\mathbb{E}\{\frac{q_M-\mu_M}{q_M}\sum_{c\in\mathcal{F}}U_{s(c)}^c(t)\mu_{s(c)b(c)}^c(t)&\\
+&\sum_{c\in\mathcal{F}}\sum_{n\in\mathcal{N}}\frac{U_n^c(t)U_{s(c)}^c(t)}{q_M}&\\
&(\sum_{j:(n,j)\in\mathcal{L}}\mu_{nj}^c(t)-\sum_{i:(i,n)\in\mathcal{L}^c}\mu_{in}^c(t))|\textbf{Q}(t)\}.&\\
\end{aligned}
\end{eqnarray}
We can rewrite the last term of RHS of (\ref{e:14}) by simple
algebra as
\begin{eqnarray}\label{e:15}
\begin{aligned}
-&\mathbb{E}\{\sum_{c\in\mathcal{F}}\sum_{(m,n)\in\mathcal{L}}\mu_{mn}^c(t)\frac{U_{s(c)}^c(t)}{q_M}(U_m^c(t)-U_n^c(t)) &\\
+& \sum_{c\in\mathcal{F}}\mu_{s(c)b(c)}^c(t)\frac{U_{s(c)}^c(t)}{q_M}(q_M-\mu_M-U_{b(c)}^c(t)) |\textbf{Q}(t)\}.&\\
\end{aligned}
\end{eqnarray}

Then, the second term and the last term of the RHS of (\ref{e:14})
are minimized by the congestion controller (\ref{eq:25}) and the
scheduling policy (\ref{eq:11}), respectively, over a set of
feasible algorithms including the stationary randomized algorithm
STAT introduced in Lemma 2 and Remark 4. 
We can substitute into the second term of RHS of
(\ref{e:14}) a stationary randomized algorithm with admitted arrival
rate vector $(r_{\epsilon,c}^*)$ and into the last term with a
stationary randomized algorithm with admitted arrival rate vector
$(r_{\epsilon,c}^*+\epsilon)$. Thus, we have:

\begin{eqnarray}\label{eq:9}
\begin{aligned}
&\Delta(t)-V\sum_{c\in\mathcal{F}}\mathbb{E}\{R_c(t)|\textbf{Q}(t)\}& \\
\leq& B
-V\sum_{c\in\mathcal{F}}r_{\epsilon,c}^*&\\
&-\sum_{c\in\mathcal{F}}\frac{U_{s(c)}^c(t)}{q_M}(\epsilon(q_M-\mu_M)-\frac{2N-1+\mu_M^2}{2})&\\
&-\sum_{c\in\mathcal{F}}(r_{\epsilon,c}^*-a_c)Z_c(t)-\sum_{c\in\mathcal{F}}(\rho_cr_{\epsilon,c}^*-Nq_M)X_c(t). &\\
\end{aligned}
\end{eqnarray}
When (\ref{eq:7}) holds, we can find $\epsilon_1>0$ such that
$\epsilon_1\leq\rho_cr_{\epsilon,c}^*-Nq_M$ $\forall
c\in\mathcal{F}$ and
$\epsilon_1\leq\frac{\epsilon(q_M-\mu_M)-\frac{2N-1+\mu_M^2}{2}}{q_M}$.
Recall that $\epsilon'$ is defined such that $r_{\epsilon,c}^*\geq
a_c+\epsilon'$ $\forall c\in\mathcal{F}$. Thus, we have:
\begin{eqnarray}\label{eq:10}
\begin{aligned}
&\Delta(t)-V\sum_{c\in\mathcal{F}}\mathbb{E}\{R_c(t)|\textbf{Q}(t)\}&\\
\leq&
B-\delta\sum_{c\in\mathcal{F}}(X_c(t)+U_{s(c)}^c(t)+Z_c(t))-V\sum_{c\in\mathcal{F}}r_{\epsilon,c}^*,&\\
\end{aligned}
\end{eqnarray}
where $\delta\triangleq\min\{\epsilon_1,\epsilon'\}$.

We take the expectation with respect to the distribution of
$\textbf{Q}$ on both sides of (\ref{eq:10}) and take the time average on
$\tau=0,...,t-1$, which leads to
\begin{eqnarray}\label{eq:13}
\begin{aligned}
&\frac{1}{t}\mathbb{E}\{L(\textbf{Q}(t))\}-\frac{V}{t}\sum_{\tau=0}^{t-1}\sum_{c\in\mathcal{F}}\mathbb{E}\{R_c(\tau)\}&\\
\leq&
B-V\sum_{c\in\mathcal{F}}r_{\epsilon,c}^*&\\
&-\frac{\delta}{t}\sum_{\tau=0}^{t-1}\sum_{c\in\mathcal{F}}\mathbb{E}\{X_c(\tau)+U_{s(c)}^c(\tau)+Z_c(\tau)\}.&\\
\end{aligned}
\end{eqnarray}

Since
$\limsup_{t\rightarrow\infty}\frac{1}{t}\sum_{\tau=0}^{t-1}\sum_c\mathbb{E}\{R_c(\tau)\}$
is bounded above (say, by a constant $B_R$ with $B_R\leq K\mu_M$)
and $\mathbb{E}\{L(\textbf{Q}(t))\}$ is nonnegative, by taking
limsup of $t$ on both sides of (\ref{eq:13}), we have:
\begin{eqnarray}\label{eq:20}
\begin{aligned}
&\limsup_{t\rightarrow\infty}\frac{1}{t}\sum_{\tau=0}^{t-1}\sum_{c\in\mathcal{F}}\mathbb{E}\{X_c(\tau)+U_{s(c)}^c(\tau)+Z_c(\tau)\}&\\
\leq&
\frac{B}{\delta}+\frac{V}{\delta}[\limsup_{t\rightarrow\infty}\frac{1}{t}\sum_{\tau=0}^{t-1}\sum_{c\in\mathcal{F}}\mathbb{E}\{R_c(\tau)\}-\sum_{c\in\mathcal{F}}r_{\epsilon,c}^*]&\\
\leq&\frac{B'}{\delta},&\\
\end{aligned}
\end{eqnarray}
which proves (\ref{eq:14}).

By taking liminf of $t$ on both sides of (\ref{eq:13}), we have
\begin{eqnarray}\label{eq:17}
\begin{aligned}
&\liminf_{t\rightarrow\infty}\frac{1}{t}\sum_{\tau=0}^{t-1}\sum_{c\in\mathcal{F}}\mathbb{E}\{R_c(\tau)\}&\\
\geq&\frac{\delta}{V}\liminf_{t\rightarrow\infty}\frac{1}{t}\sum_{\tau=0}^{t-1}\sum_{c\in\mathcal{F}}\mathbb{E}\{X_c(\tau)+U_{s(c)}^c(\tau)+Z_c(\tau)\}&\\
&-\frac{B}{V}+\sum_{c\in\mathcal{F}}r_{\epsilon,c}^*,&\\
\end{aligned}
\end{eqnarray}
which proves (\ref{eq:16}) since the first term of the RHS of
(\ref{eq:17}) is nonnegative.

\section{Further Discussions}
\subsection{Suboptimal Algorithms}
Solving MWM optimization problem can be NP-hard depending on the
underlying interference model as indicated in \cite{Lyap7}. In this
section, we introduce a group of suboptimal algorithms that aim to
achieve at least a $\gamma$ fraction of the optimal throughput. We
denote the scheduling parameters of suboptimal algorithms by
$(\mu_{mn}^{c,SUB}(t))$. For convenience, we also denote the
scheduling parameters of \emph{\textbf{ALG}} by
$(\mu_{mn}^{c,OPT}(t))$. Algorithms are called \emph{suboptimal} if
the scheduling parameters $(\mu_{mn}^{c,SUB}(t))$ satisfy the
following:
\begin{equation}\label{e:18}
\sum_{m,n}\mu_{mn}^{c_{mn}^*(t),SUB}(t)w_{mn}(t)\geq\gamma\sum_{m,n}\mu_{mn}^{c_{mn}^*(t),OPT}(t)w_{mn}(t),
\end{equation}
where $\gamma\in(0,1)$ is constant and we recall that $c_{mn}^*(t)$
and $w_{mn}(t)$ are defined in Section IV.A. In addition, the
congestion controller of suboptimal algorithms is the same as that
of \emph{\textbf{ALG}} (\ref{eq:25}).

Following the same analysis of \emph{\textbf{ALG}}, Proposition 1
holds for suboptimal algorithms, i.e., the queue backlogs are
bounded above by $q_M$, and we derive the following theorem:
\theorem{Given that
\begin{eqnarray}\label{eq:8}
\begin{aligned}
&q_M> \frac{2N-1+\mu_M^2}{2\gamma\epsilon}+\mu_M \mbox{ and }
\rho_c> \frac{Nq_M}{\gamma r_{\epsilon,c}^*} \mbox{ } \forall
c\in\mathcal{F},&\\
&\qquad\qquad\exists \epsilon_2>0 \mbox{ s.t. } \gamma
r_{\epsilon,c}^*\geq
a_c+\epsilon_2 \mbox{ } \forall c\in\mathcal{F},&\\
\end{aligned}
\end{eqnarray}
a suboptimal algorithm ensures that the virtual queues have a
time-averaged bound:
\begin{equation}\label{eq:15}
\limsup_{t\rightarrow
\infty}\frac{1}{t}\sum_{\tau=0}^{t-1}\sum_{c\in\mathcal{F}}\mathbb{E}\{U_{s(c)}^c(\tau)+X_c(\tau)+Z_c(\tau)\}\leq\frac{\bar{B}}{\delta},
\end{equation}
where $\bar{B}\triangleq B+\gamma VB_R$. In addition, a suboptimal
algorithm can achieve a throughput
\begin{equation}\label{eq:19}
\liminf_{t\rightarrow\infty}\frac{1}{t}\sum_{\tau=0}^{t-1}\sum_{c\in\mathcal{F}}\mathbb{E}\{R_c(\tau)\}\geq\gamma\sum_{c\in\mathcal{F}}r_{\epsilon,c}^*-\frac{B}{V}.
\end{equation}
}

\proof{The proof is provided in Appendix A.}\endproof

\emph{Remark 5}: From Theorem 2, given an arbitrarily small
$\epsilon$, we show that a suboptimal algorithm can \emph{at least}
achieve a throughput arbitrarily close to a fraction $\gamma$ of the
optimal results $\sum_{c\in\mathcal{F}}r_{\epsilon,c}^*$. Suboptimal
algorithms include the well-known Greedy Maximal Matching (GMM)
algorithm \cite{Lyap15} with $\gamma=\frac{1}{2}$ as well as the
solutions to the maximum weighted independent set (MWIS)
optimization problem such as GWMAX and GWMIN proposed in
\cite{graph1} with $\gamma=\frac{1}{\Delta}$, where $\Delta$ is the
maximum degree of the network topology $G$. The delay bound and
throughput tradeoff in Theorem 1 still hold in Theorem 2.

\subsection{Arbitrary Arrival Rates at Transport Layer}
Note that in the previous model description, we assumed that the
flow sources are constantly backlogged, that is, the congestion
controller (\ref{eq:25}) can always guarantee $R_c(t)=\mu_M$ when
$\frac{q_M-\mu_{M}}{q_M}U_{s(c)}^c(t)-X_c(t)\rho_c-Z_c(t)-V\leq0$.
In this subsection, we present an optimal algorithm when the flows
have arbitrary arrival rates at the transport layer.

Let $A_c(t)$ denote the arrival rate of flow $c$ packets at the
beginning of the time slot $t$ at the transport layer. We assume
that $A_c(t)$ is i.i.d. with respect to $t$ with mean $\lambda_c$.
For simplicity of analysis, we assume $(\lambda_c)$ to be in the
exterior of the capacity region $\Lambda$ so that a congestion
controller is needed and we assume that $A_c(t)$ is bounded above by
$\mu_M$ $\forall c\in\mathcal{F}$.\footnote{Note that our analysis
also works for the case when $A_c(t)$ is bounded above by some
constant $A_M$ $\forall c\in\mathcal{F}$, where $A_M\geq \mu_M$.}
Let $L_c(t)$ denote the backlog of flow $c$ data at the transport
layer which is updated as follows:
\begin{equation}\label{eq:28}
L_c(t+1)=\min\{[L_c(t)+A_c(t)-\mu_{s(c)b(c)}^c(t)]^+,L_M\},
\end{equation}
where $L_M\geq0$ is the buffer size for flow $c$ at the transport
layer. Note that we have $L_M=0$ and $L_c(t)=0$ if there is no
buffer for flow $c$ at the transport layer.

Following the idea introduced in \cite{Lyap1}, we construct a
virtual queue $Y_c(t)$ and an auxiliary variable $v_c(t)$ for each
virtual input rate $R_c(t)$, with queue dynamics for $Y_c(t)$ as
follows
\begin{equation}\label{eq:21}
Y_c(t+1)=[Y_c(t)-R_c(t)]^+ +v_c(t),
\end{equation}
where initially we have $Y_c(0)=0$. The intuition is that $v_c(t)$
serves as the function of $R_c(t)$ in congestion controller
(\ref{eq:25}) and we note that when $Y_c(t)$ is stable, we have
$r_c\geq v_c$, where $v_c$ is the time average rate for $v_c(t)$,
recalling that $r_c$ is the time average rate for $R_c(t)$. Thus,
when $Y_c(t)$ and $U_{s(c)}^c(t)$ are stable, if we can ensure the
value $\sum_cv_c$ is arbitrarily close to the optimal value
$\sum_cr_{\epsilon,c}^*$, then so is the throughput $\sum_c\mu_c$
since $\mu_c\geq r_c\geq v_c$.

Now we provide the optimal algorithm for arbitrary arrival rates at
the transport layer:

\textbf{1) Congestion Controller}:
\begin{equation}\label{eq:22}
\min_{0\leq v_c(t)\leq\mu_M} v_c(t)(\eta Y_c(t)-V),
\end{equation}
\begin{eqnarray}\label{eq:23}
\min_{R_c(t)} R_c(t)(\frac{q_M-\mu_{M}}{q_M}U_{s(c)}^c(t)-\eta
Y_c(t)-X_c(t)\rho_c-Z_c(t))
\end{eqnarray}
\begin{displaymath}
\mbox{s.t. }\qquad 0\leq R_c(t)\leq\min\{L_c(t)+A_c(t),\mu_M\}
\end{displaymath}
where $\eta>0$ is a weight associated with the virtual queue
$Y_c(t)$. Note that (\ref{eq:22}) and (\ref{eq:23}) can be solved
independently. Specifically, when $\eta Y_c(t)-V\geq0$, $v_c(t)$ is
set to zero; Otherwise, $v_c(t)=\mu_M$. When
$\frac{q_M-\mu_{M}}{q_M}U_{s(c)}^c(t)-\eta
Y_c(t)-X_c(t)\rho_c-Z_c(t)\geq0$, $R_c(t)$ is set to zero;
Otherwise, $R_c(t)=\min\{L_c(t)+A_c(t),\mu_M\}$.

\textbf{2) Scheduling Policy}: The scheduling algorithm is the same
as that of \emph{\textbf{ALG}} provided in Section IV.B, except for
the updated constraints:
$0\leq\mu_{s(c)b(c)}^c(t)\leq\min\{L_c(t)+A_c(t),\mu_M\}$.

Since the scheduling policy is not changed, Proposition 1 still
holds. And we present a theorem below for the performance of the
algorithm: \theorem{Given that
\begin{displaymath}
q_M> \frac{2N-1+\mu_M^2}{2\epsilon}+\mu_M \mbox{ and } \rho_c>
\frac{Nq_M}{r_{\epsilon,c}^*} \mbox{ } \forall c\in\mathcal{F},
\end{displaymath}
the algorithm ensures that the virtual queues have a time-averaged
bound:
\begin{displaymath}
\limsup_{t\rightarrow
\infty}\frac{1}{t}\sum_{\tau=0}^{t-1}\sum_{c\in\mathcal{F}}\mathbb{E}\{U_{s(c)}^c(\tau)+X_c(\tau)+Z_c(\tau)+Y_c(\tau)\}\leq\frac{B_2}{\delta'},
\end{displaymath}
where $B_2\triangleq B+K\eta\mu_M^2+VB_R$ and $\delta'$ is constant
positive number. In addition, the algorithm can achieve a throughput
\begin{displaymath}
\liminf_{t\rightarrow\infty}\frac{1}{t}\sum_{\tau=0}^{t-1}\sum_{c\in\mathcal{F}}\mathbb{E}\{v_c(\tau)\}\geq\sum_{c\in\mathcal{F}}r_{\epsilon,c}^*-\frac{B_1}{V},
\end{displaymath}
where $B_1\triangleq B+K\eta\mu_M^2$.}

\proof{The proof is provided in Appendix B.}\endproof

Theorem 3 shows that optimality is preserved and
$O(\frac{1}{\epsilon})$ delay scaling is kept.

\subsection{Employing Delayed Queue Backlog Information}
Recall that in \emph{\textbf{ALG}}, congestion controller
(\ref{eq:25}) is performed at the transport layer and link weight
assignment in (\ref{e:6}) is performed locally at each link. Thus,
in order to account for the propagation delay of queue information,
we employ delayed queue backlog of $(X_c(t))$ in (\ref{eq:25}) and
employ delayed queue backlog of $(U_{s(c)}^c(t))$ for links in
$\mathcal{L}$ in (\ref{e:6}). Specifically, we rewrite (\ref{eq:25})
in \emph{\textbf{ALG}} as:
\begin{equation}\label{e:7}
\min
R_c(t)(\frac{(q_M-\mu_{M})U_{s(c)}^c(t)}{q_M}-X_c(t-T)\rho_c-Z_c(t)-V),
\end{equation}
where $T$ is an integer number that is larger than the maximum
propagation delay from a source to a node, and we rewrite
(\ref{e:6}) as:
\begin{eqnarray}\label{e:8}
w_{mn}^c(t)=\left\{\begin{aligned} &
\frac{U_{s(c)}^c(t-T)}{q_M}[U_m^c(t)-U_n^c(t)],&\\
&\qquad\qquad\qquad\mbox{ if }
(m,n)\in \mathcal{L}, &\\
& \frac{U_{s(c)}^c(t)}{q_M}[q_M-\mu_M-U_{b(c)}^c(t)],&\\
&\qquad\qquad\qquad\mbox{ if }
(m,n)=(s(c),b(c)), &\\
&0, \qquad\qquad\quad\mbox{ otherwise. }&\\
\end{aligned} \right.
\end{eqnarray}

Proposition 1 still holds and we present a theorem for the
scheduling algorithm using delayed queue backlog information, which
maintains the throughput optimality and $O(\frac{1}{\epsilon})$
scaling in delay bound: \theorem{Given that
\begin{displaymath}
q_M> \frac{2N-1+\mu_M^2}{2\epsilon}+\mu_M \mbox{ and } \rho_c>
\frac{Nq_M}{r_{\epsilon,c}^*} \mbox{ } \forall c\in\mathcal{F},
\end{displaymath}
the algorithm ensures that the virtual queues have a time-averaged
bound:
\begin{displaymath}
\limsup_{t\rightarrow
\infty}\frac{1}{t}\sum_{\tau=0}^{t-1}\sum_{c\in\mathcal{F}}\mathbb{E}\{U_{s(c)}^c(\tau)+X_c(\tau)+Z_c(\tau)\}\leq\frac{B_4}{\delta},
\end{displaymath}
where $B_4\triangleq B_3+VB_R$ and $B_3\triangleq
B+KN\mu_MT+Nq_MT\mu_M\rho_c+K\rho_c^2\mu_M^2T$. In addition, the
algorithm can achieve a throughput
\begin{displaymath}
\liminf_{t\rightarrow\infty}\frac{1}{t}\sum_{\tau=0}^{t-1}\sum_{c\in\mathcal{F}}\mathbb{E}\{R_c(\tau)\}\geq\sum_{c\in\mathcal{F}}r_{\epsilon,c}^*-\frac{B_3}{V}.
\end{displaymath}
}

\proof{The proof is provided in Appendix C.}\endproof

On employing delayed queue backlogs, we can extend the centralized
optimization problem (\ref{eq:11}) to distributed implementations
with methods introduced in Remark 3. 

\subsection{Arbitrary Link Capacities and Arbitrary Interference Models with Fading Channels}
Recall that in the model description in Section III, the link
capacity is assumed constant (one packet per slot) and
node-exclusive model is employed. In this subsection, we extend the
model to arbitrary link capacities and arbitrary interference models
with fading channels of finite channel states. Thus, instead of
(\ref{e:4}), we have $(\mu_{mn}^c(t))_{(m,n)\in\mathcal{L}}\in
I(t)$, where $I(t)$ is the feasible activation set for time slot $t$
determined by the underlying interference model and current channel
states, with link capacity constraints
$\sum_{c\in\mathcal{F}}\mu_{mn}^c(t)\leq l_{mn}$, where $l_{mn}$ is
the arbitrarily chosen link capacity for a link
$(m,n)\in\mathcal{L}$. We define $l_n\triangleq
\max_{(\mu_{mn}^c(t))\in
I(t)}\sum_{c\in\mathcal{F}}\sum_{m:(m,n)\in\mathcal{L}}\mu_{mn}^c(t)$.
Note that it is clear that $l_n\leq
\sum_{m:(m,n)\in\mathcal{L}}l_{mn}$. Then we can update the
optimization (\ref{eq:11}) and weight assignment (\ref{e:6}),
respectively, as follows:
\begin{displaymath}
\begin{aligned}
&\max_{(\mu_{mn}^c(t))}\sum_{m,n}\mu_{mn}^{c_{mn}^*(t)}(t)w_{mn}(t)&\\
\mbox{s.t. }&(\mu_{mn}^c(t))_{(m,n)\in\mathcal{L}}\in I(t) \mbox{
and }\mu_{s(c)b(c)}(t)\leq\mu_M \mbox{ }\forall c\in\mathcal{L}.&
\end{aligned}
\end{displaymath}
\begin{displaymath}
w_{mn}^c(t)=\left\{\begin{aligned} &
\frac{U_{s(c)}^c(t)}{q_M}[U_m^c(t)-U_n^c(t)-l_n],\mbox{ if }
(m,n)\in \mathcal{L}, &\\
& \frac{U_{s(c)}^c(t)}{q_M}[q_M-\mu_M-U_{b(c)}^c(t)],&\\
&\qquad\qquad\qquad\mbox{ if }
(m,n)=(s(c),b(c)), &\\
&0, \qquad\qquad\quad\mbox{ otherwise. }&\\
\end{aligned} \right.
\end{displaymath}

It is not difficult to check that Proposition 1 still holds with
$q_M\geq \max\{\max_{n\in\mathcal{N}}l_n,\mu_M\}$ and Theorem 1
holds with a different definition of constant $B$. The above
modified algorithm can be further extended to solve power allocation
problems, where we refer interested readers to our recent work
\cite{ICC11}.

\section{Numerical Results}
\begin{figure}[htbp]
\centering
{\includegraphics[height=8.5cm,width=0.43\textwidth]{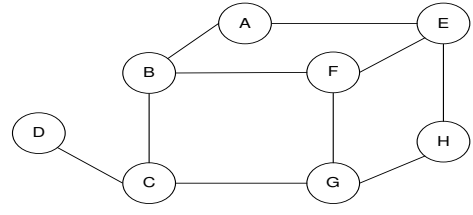}}
\vspace{-50mm} \caption{Network topology for simulations}
\label{fig1} 
\end{figure}

In this section, we present the simulation results for the proposed
optimal algorithm \emph{\textbf{ALG}}. Simulations are run in Matlab
2009A with results averaged over $10^5$ time slots. In the network
topology illustrated in Figure \ref{fig1}, there are three
source-destination pairs $(A,G)$, $(D,E)$ and $(F,H)$ with same
Poisson arrival rates and $\mu_M=2$. The required minimum data rate
for the three flows are all set to $0.1$. We denote by \emph{BP} the
back-pressure scheduling algorithm in \cite{Lyap0} with a congestion
controller in \cite{Lyap1}, and denote by \emph{Finite Buffer} the
cross-layer algorithm developed in \cite{Lyap14} with buffer size
equal to the queue length limit $q_M$. Note that it is shown in
simulation results in \cite{Lyap14} that Finite Buffer algorithm
ensures much smaller internal queue length (of nodes excluding the
source node) than BP algorithm (and queue length is related to delay
performance). We set the control parameter $V=1000$, where in
simulations we find that a higher $V$ cannot further improve the
throughput.

\begin{table*}[htbp]
\renewcommand{\arraystretch}{1.3}
\begin{center}
\caption{Throughput performance of \emph{\textbf{ALG}} when sources
are backlogged at the transport layer} \label{tab1}
\begin{tabular}{|c|c|c|c|c|c|}
\hline \hline
${ }$ & \textbf{\emph{ALG}} ($\rho_c=150$) & \textbf{\emph{ALG}} ($\rho_c=300$) & \textbf{\emph{ALG}} ($\rho_c=3000$) & \textbf{\emph{ALG}} ($\rho_c=30000$) & BP\\
\hline
Throughput (sum for three flows)  & $0.9368$ & $1.1834$ & $1.2007$ & $1.2305$ & $1.2315$\\
\hline
End-to-end delay (averaged over three flows) & $45.76$ & $131.47$ & $1.514\times10^3$ & $1.3687\times10^4$ & $3.753\times10^4$\\
\hline \hline
\end{tabular}
\end{center}
\end{table*}

We first illustrate in Table \ref{tab1} the throughput optimality of
\emph{\textbf{ALG}} when the sources are constantly backlogged. We
loosen the delay constraint as $\rho_c=30q_M$. As we increase the
control parameter $q_M$, the \emph{\textbf{ALG}} achieves a
throughput approaching the throughput of BP algorithm which is known
to be optimal. We also note that this approximation in throughput
results in worse average end-to-end delay performance, which
complies with Remark 1.

\begin{figure}[htbp]
\centering
{\includegraphics[height=5.5cm,width=0.54\textwidth]{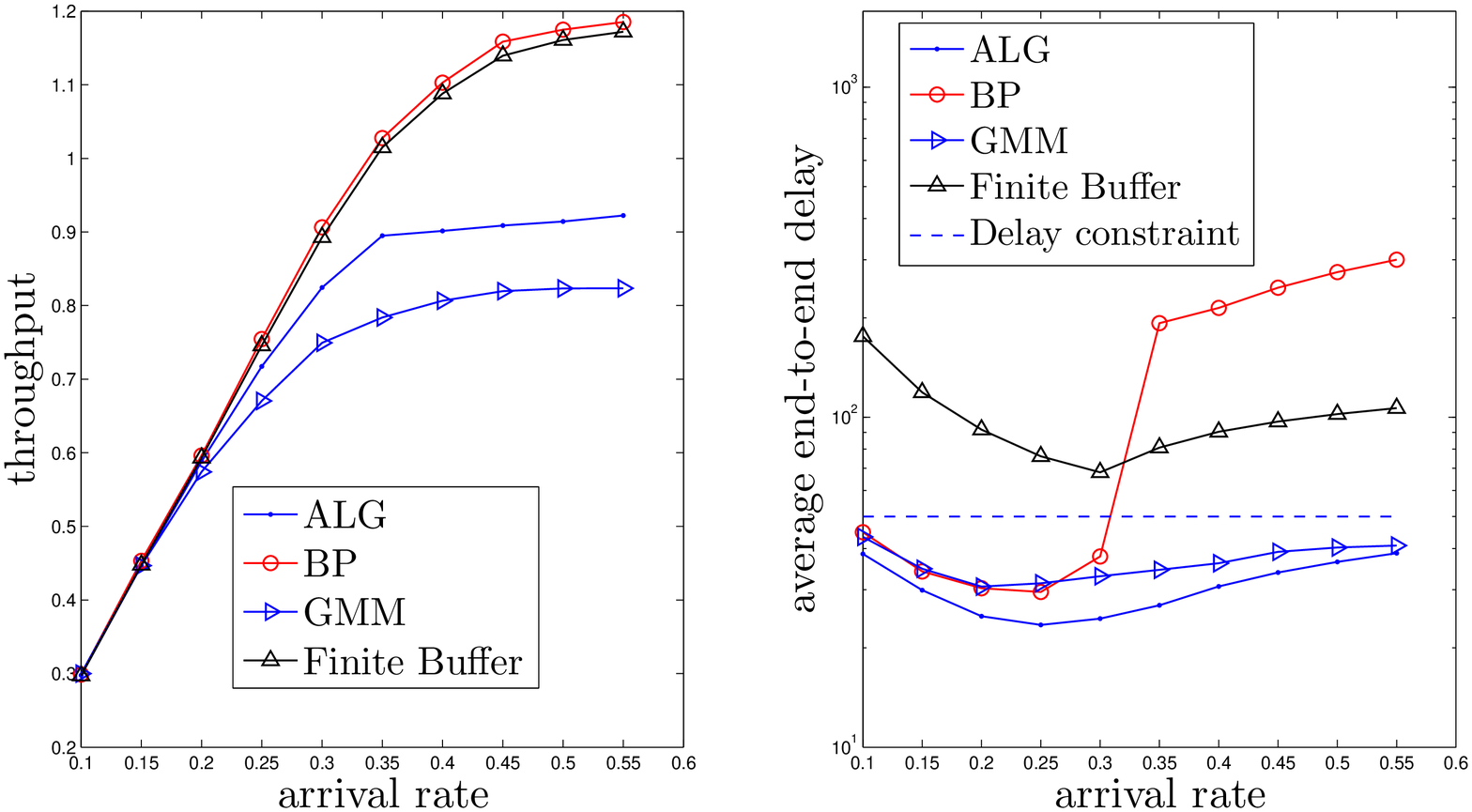}}
\caption{Throughput and delay tradeoff under Alg. with performances
compared to Finite Buffer algorithm and BP algorithm, with varying
arrival rates at the transport layer.} \label{fig2} 
\end{figure}

We then illustrate the throughput and delay tradeoff for both the
\textbf{\emph{ALG}} and its corresponding suboptimal GMM algorithm
in Figure \ref{fig2} for the case of arbitrary arrival rates at
transport layer with $L_M=0$, where we set $q_M=5$ and $\rho_c=50$
for each flow $c$. Note that this pair of $q_M$ and $\rho_c$ shows
that the bound in (\ref{eq:7}) is actually quite loose, and thus our
algorithm can achieve better delay performance than stated in
(\ref{eq:7}). Figure \ref{fig2} shows that the average end-to-end
delay under \emph{\textbf{ALG}} is well below the constraint
($\rho_c=50$) and lower than that under BP and Finite Buffer
algorithms. The throughput of \emph{\textbf{ALG}} is close to
(although lower than) that of the optimal BP algorithm when arrival
rates are small ($\leq0.3$). Specifically, when the arrival rate is
$0.3$, \emph{\textbf{ALG}} achieves a throughput $10\%$ more than
the GMM algorithm and $9.0\%$ less than BP algorithm, with an
average end-to-end delay $35.2\%$ less than the BP algorithm. In the
large-input-rate-region ($>0.3$), we also observe that the delay in
both the BP and Finite Buffer algorithm violates the delay
constraints. In addition, in the above illustrated scenarios with
backlogged and arbitrary arrival rates, the minimum arrival rates
and average end-to-end delay requirements are satisfied for
\emph{individual} flows under \emph{\textbf{ALG}}. As a side note,
the average end-to-end delay in all four algorithms in Figure
\ref{fig2} first decreases, which can be explained by the intuition
that all the algorithms are based on back-pressure of links (i.e.,
queue backlog difference of links) and the queue backlog difference
tends to be larger for each hop with a larger arrival rate. When
arrival rate further increases, congestion level becomes higher
since more packets are admitted into the network.

\section{Conclusions and Future works}
In this paper, we proposed a cross-layer framework which approaches
the optimal throughput arbitrarily close for a general multi-hop
wireless network. We show a tradeoff between the throughput and
average end-to-end delay bound while satisfying the minimum data
rate requirements for individual flows.

Our work aims at a better understanding of the fundamental
properties and performance limits of QoS-constrained multi-hop
wireless networks. While we show an $O(\frac{1}{\epsilon})$ delay
bound with $\epsilon$-loss in throughput, how small the actual delay
can become still remains elusive, which is dependent on specific
network topologies. In our future work, we will investigate the
capacity region under end-to-end delay constraints applied to
different network topologies. Our future work will also involve
power management in the scheduling policies.

\appendices
\section{Proof of Theorem 2}
\proof{Let $\Delta^{SUB}(t)$ denote the corresponding Lyapunov drift
of a suboptimal algorithm which takes the same form as
(\ref{eq:12}). By analyzing (\ref{e:14})(\ref{e:15}) which also hold
for suboptimal algorithms, we note that the second term of RHS of
(\ref{e:14}) is always non-positive ensured by the congestion
controller (\ref{eq:25}). Employing (\ref{e:18}) to
(\ref{e:14})(\ref{e:15}), we derive the following

\begin{eqnarray}\label{e:17}
\begin{aligned}
&\Delta^{SUB}(t)-V\sum_{c\in\mathcal{F}}\mathbb{E}\{R_c(t)|\textbf{Q}(t)\}& \\
\leq&B+\gamma\sum_{c\in\mathcal{F}}\mathbb{E}\{R_c(t)(\frac{(q_M-\mu_{M})U_{s(c)}^c(t)}{q_M}&\\
&\qquad\qquad\qquad-X_c(t)\rho_c-Z_c(t)-V)|\textbf{Q}(t)\}&\\
+&Nq_M\sum_{c\in\mathcal{F}}X_c(t)+\sum_{c\in\mathcal{F}}a_cZ_c(t)&\\
+&\frac{1}{2}\sum_{c\in\mathcal{F}}\frac{(2N-1+\mu_M^2)U_{s(c)}^c(t)}{q_M} &\\
-&\gamma\mathbb{E}\{\frac{q_M-\mu_M}{q_M}\sum_{c\in\mathcal{F}}U_{s(c)}^c(t)\mu_{s(c)b(c)}^{c,SUB}(t)&\\
+&\sum_{c\in\mathcal{F}}\sum_{n\in\mathcal{N}}\frac{U_n^c(t)U_{s(c)}^c(t)}{q_M}&\\
&(\sum_{j}\mu_{nj}^{c,SUB}(t)-\sum_{i}\mu_{in}^{c,SUB}(t))|\textbf{Q}(t)\},&\\
\end{aligned}
\end{eqnarray}

Following the steps in proving (\ref{eq:9}), we have from
(\ref{e:17})
\begin{eqnarray}\label{e:19}
\begin{aligned}
&\Delta(t)^{SUB}-V\sum_{c\in\mathcal{F}}\mathbb{E}\{R_c(t)|\textbf{Q}(t)\}& \\
\leq& B
-V\gamma\sum_{c\in\mathcal{F}}r_{\epsilon,c}^*&\\
-&\sum_{c\in\mathcal{F}}\frac{U_{s(c)}^c(t)}{q_M}(\gamma\epsilon(q_M-\mu_M)-\frac{2N-1+\mu_M^2}{2})&\\
-&\sum_{c\in\mathcal{F}}(\gamma r_{\epsilon,c}^*-a_c)Z_c(t)-\sum_{c\in\mathcal{F}}(\gamma\rho_cr_{\epsilon,c}^*-Nq_M)X_c(t). &\\
\end{aligned}
\end{eqnarray}
Employing the conditions (\ref{eq:8}) and following the steps in
proving (\ref{eq:20}) and (\ref{eq:17}), we can prove Theorem 2.}
\endproof

\section{Proof of Theorem 3}
Before we proceed to the proof, we extend the stationary randomized
algorithm STAT introduced in Lemma 2 and Remark 4. Given
$(\theta_c)$ introduced in Lemma 2 and given flow $c$ at node $n$,
recall that $(A_c(t))$ is i.i.d. with mean $(\lambda_c)$ and
$(\lambda_c)>(\theta_c)$ element-wise. The flow control for STAT can
be given as: Admit $\mu_{s(c)b(c)}^{c,STAT}(t)=A_c(t)$ w.p.
$\frac{\theta_c}{\lambda_c}$; otherwise,
$\mu_{s(c)b(c)}^{c,STAT}(t)=0$. Then
$\mathbb{E}\{\mu_{s(c)b(c)}^{c,STAT}(t)\}=\theta_c$, $\forall t$.
Now take $v_c^{STAT}(t)=R_c^{STAT}(t)=\mu_{s(c)b(c)}^{c,STAT}(t)$
$\forall c\in\mathcal{F}$. Then we also have
$\mathbb{E}\{v_c^{STAT}(t)\}=\mathbb{E}\{R_c^{STAT}(t)\}=\theta_c$.
Note that $R_c^{STAT}(t)\leq A_c(t)\leq\min\{L_c(t)+A_c(t),\mu_M\}$
and $v_c^{STAT}(t)\leq \mu_M$.

Now we present the proof. \proof{We define the Lyapunov function as
$L(\textbf{Q}'(t))=L(\textbf{Q}(t))+\frac{\eta}{2}\sum_{c\in\mathcal{F}}Y_c^2(t)$
and the Lyapunov drift as $\Delta'(t)=\mathbb{E}\{L(
\textbf{Q}'(t+1)) -L(\textbf{Q}'(t)) |\textbf{Q}'(t)\}$, where
$\textbf{Q}'(t)=(\textbf{Q}(t),(Y_c(t)))$. From the virtual queue
dynamics (\ref{eq:21}) and Lemma 1, we have
\begin{eqnarray}\label{eq:24}
\begin{aligned}
&\frac{\eta}{2}\sum_{c\in\mathcal{F}}(Y_c(t+1)^2-Y_c(t)^2)&\\
\leq&\frac{\eta}{2}\sum_{c\in\mathcal{F}}(R_c(t)^2+v_c(t)^2-2Y_c(t)(R_c(t)-v_c(t)))&\\
\leq&K\eta\mu_M^2-\sum_{c\in\mathcal{F}}\eta Y_c(t)(R_c(t)-v_c(t)).  &\\
\end{aligned}
\end{eqnarray}

Following the steps in deriving (\ref{e:14})(\ref{e:15}), we have
\begin{eqnarray}\label{eq:27}
\begin{aligned}
&\Delta'(t)-V\sum_{c\in\mathcal{F}}\mathbb{E}\{v_c(t)|\textbf{Q}'(t)\}& \\
\leq&B_1+\sum_{c\in\mathcal{F}}\mathbb{E}\{v_c(t)(\eta
Y_c(t)-V)|\textbf{Q}'(t)\}&\\
+&\sum_{c\in\mathcal{F}}\mathbb{E}\{R_c(t)(\frac{(q_M-\mu_{M})U_{s(c)}^c(t)}{q_M}&\\
&\qquad\qquad\quad-\eta Y_c(t)-X_c(t)\rho_c-Z_c(t))|\textbf{Q}'(t)\}&\\
+&Nq_M\sum_{c\in\mathcal{F}}X_c(t)+\sum_{c\in\mathcal{F}}a_cZ_c(t)&\\
+&\frac{1}{2}\sum_{c\in\mathcal{F}}\frac{(2N-1+\mu_M^2)U_{s(c)}^c(t)}{q_M} &\\
-&\mathbb{E}\{\sum_{c\in\mathcal{F}}\sum_{(m,n)\in\mathcal{L}}\mu_{mn}^c(t)\frac{U_{s(c)}^c(t)}{q_M}(U_m^c(t)-U_n^c(t)) &\\
+& \sum_{c\in\mathcal{F}}\mu_{s(c)b(c)}^c(t)\frac{U_{s(c)}^c(t)}{q_M}(q_M-\mu_M-U_{b(c)}^c(t)) |\textbf{Q}'(t)\},&\\
\end{aligned}
\end{eqnarray}

The second term, third term and the last term of the RHS of
(\ref{eq:27}) are minimized by the congestion controller
(\ref{eq:22}), (\ref{eq:23}) and the scheduling policy
(\ref{eq:11}), respectively, over a set of feasible algorithms
including the stationary randomized algorithm STAT. Substitute into
the second term of RHS of (\ref{eq:27}) a stationary randomized
algorithm with admitted arrival rate vector
$(r_{\epsilon,c}^*-\frac{1}{2}\epsilon')$, the third term a
stationary randomized algorithm with admitted arrival rate vector
$(r_{\epsilon,c}^*)$ and the last term a stationary randomized
algorithm with admitted arrival rate vector
$(r_{\epsilon,c}^*+\epsilon)$. Then, following the steps in proving
Theorem 1, we can prove Theorem 3.}
\endproof

\section{Proof of Theorem 4}
\proof{According to queue dynamics (\ref{eq:4})(\ref{eq:2}), we
obtain
\begin{eqnarray}\label{e:9}
\begin{aligned}
&U_{s(c)}^c(t)-\mu_MT\leq U_{s(c)}^c(t-T)\leq U_{s(c)}^c(t)+\mu_MT,&\\
&X_c(t)-Nq_MT\leq X_c(t-T)\leq X_c(t)+\rho_c\mu_MT.&\\
\end{aligned}
\end{eqnarray}
Employing the above inequalities to (\ref{e:14})(\ref{e:15}), we
have
\begin{displaymath}
\begin{aligned}
&\Delta(t)-V\sum_{c\in\mathcal{F}}\mathbb{E}\{R_c(t)|\textbf{Q}(t)\}& \\
\leq&B+\sum_{c\in\mathcal{F}}\mathbb{E}\{R_c(t)(\frac{(q_M-\mu_{M})U_{s(c)}^c(t)}{q_M}&\\
&\qquad\qquad\qquad-X_c(t-T)\rho_c-Z_c(t)-V)|\textbf{Q}(t)\}&\\
+&Nq_M\sum_{c\in\mathcal{F}}X_c(t)+\sum_{c\in\mathcal{F}}a_cZ_c(t)+K\rho_c^2\mu_M^2T+\frac{1}{2}KN\mu_MT&\\
+&\frac{1}{2}\sum_{c\in\mathcal{F}}\frac{(2N-1+\mu_M^2)U_{s(c)}^c(t)}{q_M} &\\
-&\mathbb{E}\{\sum_{c\in\mathcal{F}}\sum_{(m,n)\in\mathcal{L}}\mu_{mn}^c(t)\frac{U_{s(c)}^c(t-T)}{q_M}(U_m^c(t)-U_n^c(t)) &\\
+& \sum_{c\in\mathcal{F}}\mu_{s(c)b(c)}^c(t)\frac{U_{s(c)}^c(t)}{q_M}(q_M-\mu_M-U_{b(c)}^c(t)) |\textbf{Q}(t)\}.&\\
\end{aligned}
\end{displaymath}
The second term and the last term of the RHS of the above inequality
are minimized by the congestion controller (\ref{e:7}) and the
scheduling policy (\ref{eq:11}) with weight assignment (\ref{e:8}),
respectively, over a set of feasible algorithms including the
stationary randomized algorithm STAT. Substitute into the second
term of RHS a stationary randomized algorithm with admitted arrival
rate vector $(r_{\epsilon,c}^*)$ and the last term a stationary
randomized algorithm with admitted arrival rate vector
$(r_{\epsilon,c}^*+\epsilon)$. Then, employing the inequalities
(\ref{e:9}) and following the steps in proving Theorem 1, we can
prove Theorem 4. \endproof}

\end{document}